\documentclass[aps,prx,superscriptaddress,twocolumn,amsmath]{revtex4-2}

\usepackage[SquareTraceBrackets]{quantum}
\usepackage{graphicx,natbib,mathrsfs,amsmath,amsbsy,bm,optidef}
\usepackage{optidef,siunitx}
\usepackage[normalem]{ulem}
\usepackage[caption=false]{subfig}
\usepackage{nameref,varioref,tabularx}
\usepackage[dvipsnames]{xcolor}
\definecolor{myblue}{named}{MidnightBlue}
\DeclareGraphicsExtensions{.pdf, .png}
 
\usepackage[colorlinks=true,citecolor=myblue,linkcolor=myblue,urlcolor=myblue]{hyperref}
\usepackage{cleveref}

\makeatletter
\def\thickhline{%
  \noalign{\ifnum0=`}\fi\hrule \@height \thickarrayrulewidth \futurelet
   \reserved@a\@xthickhline}
\def\@xthickhline{\ifx\reserved@a\thickhline
               \vskip\doublerulesep
               \vskip-\thickarrayrulewidth
             \fi
      \ifnum0=`{\fi}}
\makeatother

\begin{document}
\title{Finite resource performance of small satellite-based quantum key distribution missions}

\author{Tanvirul Islam}
\email{cqtmti@nus.edu.sg}
\thanks{These authors contributed equally}
\affiliation{Centre for Quantum Technologies, National University of Singapore, 3 Science Drive 2, 117543 Singapore}
\author{Jasminder S. Sidhu}
\email{jsmdrsidhu@gmail.com}
\thanks{These authors contributed equally}
\affiliation{SUPA Department of Physics, University of Strathclyde, Glasgow, G4 0NG, UK}
\author{Brendon L. Higgins}
\email{brendon.higgins@uwaterloo.ca}
\affiliation{Institute for Quantum Computing and Department of Physics and Astronomy, University of Waterloo, Waterloo, ON N2L 3G1, Canada}
\author{Thomas Brougham}
\affiliation{SUPA Department of Physics, University of Strathclyde, Glasgow, G4 0NG, UK}
\author{Tom Vergoossen}
\affiliation{SpeQtral Pte.\ Ltd., 73 Science Park Drive Science Park 1, 118254 Singapore}
\author{Daniel K. L. Oi}
\affiliation{SUPA Department of Physics, University of Strathclyde, Glasgow, G4 0NG, UK}
\author{Thomas Jennewein}
\affiliation{Institute for Quantum Computing and Department of Physics and Astronomy, University of Waterloo, Waterloo, ON N2L 3G1, Canada}
\author{Alexander Ling}
\affiliation{Centre for Quantum Technologies, National University of Singapore, 3 Science Drive 2, 117543 Singapore}
\affiliation{Department of Physics, National University of Singapore, Blk S12, 2 Science Drive 3, 117551 Singapore}

\begin{abstract}
    In satellite-based quantum key distribution (QKD), the number of secret bits that can be generated in a single satellite pass over the ground station is severely restricted by the pass duration and the free-space optical channel loss. High channel loss may decrease the signal-to-noise ratio due to background noise, reduce the number of generated raw key bits, and increase the quantum bit error rate (QBER), all of which have detrimental effects on the output secret key length. Under finite-size security analysis, higher QBER increases the minimum raw key length necessary for non-zero secret key length extraction due to less efficient reconciliation and post-processing overheads. We show that recent developments in finite key analysis allow three different small-satellite-based QKD projects CQT-Sat, UK-QUARC-ROKS, and QEYSSat to produce secret keys even under very high loss conditions, improving on estimates based on previous finite key bounds. This suggests that satellites in low Earth orbit can satisfy finite-size security requirements, but remains challenging for satellites further from Earth. We analyse the performance of each mission to provide an informed route toward improving the performance of small-satellite QKD missions. We highlight the short and long-term perspectives on the challenges and potential future developments in small-satellite-based QKD and quantum networks. In particular, we discuss some of the experimental and theoretical bottlenecks, and improvements necessary to achieve QKD and wider quantum networking capabilities in daylight and at different altitudes.    
\end{abstract}

\maketitle

\section{Introduction}
\label{introduction}

\noindent
The emergence of terrestrial quantum networks in large metropolitan areas demonstrates an increasing maturity of quantum technologies. A networked infrastructure enables increased capabilities for distributed applications in delegated quantum computing~\cite{Donkor2004distributed, Meter2016the}, quantum communications~\cite{Liorni2021, Wallnofer2021}, and quantum sensing~\cite{Sidhu2020_AVS}. However, extending these applications over global scales is currently not possible owing to exponential losses in optical fibres. Space-based segments provide a practical route to overcome this and realise global quantum networking~\cite{Gundogan2021npjQI, gundogan2021topical, Gundogan2023time}. Satellite-based Quantum Key Distribution (SatQKD) has become a precursor to long-range applications of general quantum communication~\cite{Sidhu2021advances, belenchia2021quantum}. Although a general-purpose quantum network~\cite{wehner2018quantum} will require substantial advancements in quantum memories and routing techniques, a satellite-based QKD system adds to the progress of global-scale quantum networks by driving the maturation of space-based long-distance quantum links.

There has been growing interest in satellite-based quantum key distribution. The recent milestone achievements by the Micius satellite~\cite{jianwei2018progress} which demonstrated space-to-ground QKD and entanglement distribution have energized this interest. Micius, being a relatively large satellite, leaves open the question of using smaller satellites to perform satellite-based QKD---there have been feasibility studies for small-satellite-based QKD and CubeSat-based pathfinder missions~\cite{villar2020entanglement} for QKD applications. 

The recent surge in efforts emphasizes the importance of understanding specific limitations to the performance of different SatQKD systems. For low-Earth orbit (LEO) satellites, a particular challenge is the limited time window to establish and maintain a quantum channel with an optical ground station (OGS). This limitation disproportionately constrains the volume of secure keys that can be generated due to a pronounced impact of statistical fluctuations in estimated parameters~\cite{Sidhu2023finite, Sidhu2023satellite}.

In this work, we give a scientific perspective on the progress of small satellite-based quantum key distribution under resource constraints. More specifically, we analyze three different mission configurations: the Singapore Centre for Quantum Technologies' CQT-Sat, the UK Quantum Research CubeSat/Responsive Operations for Key Services (QUARC/ROKS) satellite, and the Canadian Quantum Encryption and Science Satellite (QEYSSat) on which the authors are actively participating. In addition, these three missions are representative of near-term small satellite-based QKD missions.

\begin{figure*}[t!]
    \centering
    \includegraphics[width=0.87\textwidth]{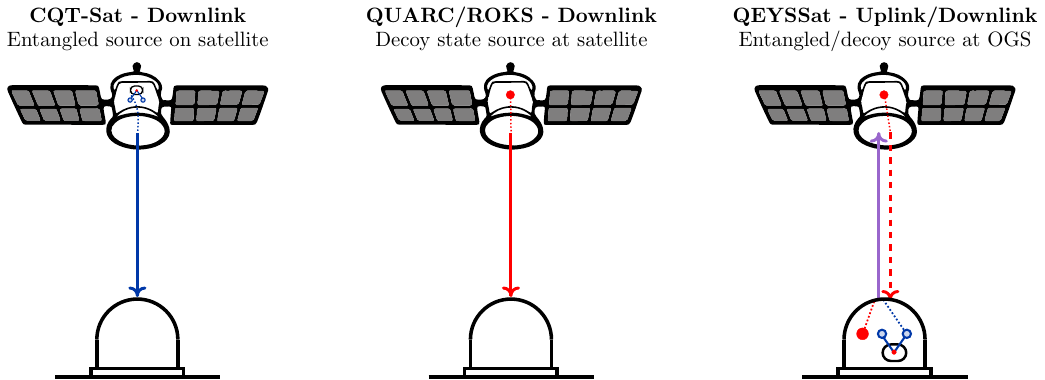}
    \caption{Quantum channel configuration for three different SatQKD missions. Each mission implements a different combination of QKD protocols and quantum channel configurations between an OGS and an orbiting satellite. The Singaporean CQT-Sat mission (left) implements the entanglement-based BBM92 protocol (blue arrow) in a downlink configuration. For this mission, one of the photon pairs is measured on board and the other is transmitted to the OGS. The UK QUARC/ROKS mission (middle) implements the weak coherent pulse (WCP) decoy-state BB84 protocol (red arrow) in a downlink configuration. The Canadian QEYSSat mission (right), implements both the decoy BB84 and BBM92 protocols (purple arrow) in an uplink configuration and intends to also incorporate a decoy BB84 downlink.}
    \label{fig:mission_configurations}
\end{figure*}%

The quantum channel configuration for each mission is illustrated in Figure~\ref{fig:mission_configurations}. With the exception of a downlink entanglement-based channel, these configurations cover uplink and downlink with entanglement-based and prepare-and-measure based QKD. These configurations give representative quantum channels to support the capability of a range of distributed applications. Depending on the ground station location and the specific LEO orbit, a satellite may have a limited number of passes over the OGS for which QKD key generation is possible---for example, current technology requires that passes are conducted during nighttime. Therefore, it is important to understand the conditions that allow a SatQKD system to produce secret keys successfully from a single pass over the OGS. More specifically, for any given satellite overpass, how many secret key bits can be generated? We answer this for the three mission configurations by revisiting the supporting theory and modelling of key generation~\cite{Lim2014_PRA, Tomamichel2017_QIP, Zhang2017_PRA, Yin2020_SR, lim2021security}.  It is shown that all three missions demonstrate enhanced key generation with the latest advancements in finite key analysis. We conclude by looking at the prospects for satellites at higher altitudes where the longer access time for a ground receiver does not overcome the increased diffraction loss.

Based on the performance analysis of each of these missions, we provide a broad and future-looking perspective for global quantum communications with a specific outlook on the outstanding challenges for SatQKD and long-term perspectives. In particular, we explore how improved finite-key analyses can improve SatQKD performances and more widely how advances in hardware can support greater capabilities for networked quantum technologies. This perspective provides a view of the medium to long-term challenges and milestones that present building blocks for enabling the quantum internet.


\section {Satellite-to-ground entanglement-based QKD using BBM92}

\noindent
CQT-Sat is a concept for 12U nano-satellite capable of performing space-to-ground entanglement-based QKD. Its precursor SpooQy-1, demonstrated~\cite{villar2020entanglement} successful launch and operation of a miniaturized polarization-entangled photon pair source in LEO. The subsequent instruments will build upon this to perform space-to-ground entanglement distribution and demonstrate entanglement-based BBM92 QKD.

During a satellite's overpass over the ground station, the link loss for the downlink quantum channel will depend on the relative distance between the satellite and the OGS. Using a variable attenuator, a tabletop setup can emulate a time-varying satellite-to-ground link loss (a similar experiment was conducted previously in the context of QEYSSat~\cite{Bourgoin2015b}). This enables us an estimation of the achievable raw key length and overall QBER for various satellite passes. Using these parameters we perform finite key analysis and show that CQT-Sat can successfully generate shared secret keys between the satellite and OGS when the maximum elevation is as low as $33^\circ$.    


\subsection{System configuration}
\noindent
The satellite quantum source generates polarization-entangled photon pairs by superposing orthogonally polarized photons created from spontaneous parametric down-conversion using two pump decay paths~\cite{anwar2021entangled}. Detailed design of a functional model of the source and associated design trade-offs can be found in~\cite{perumangatt2021realizing}. The source generates pairs of polarization-entangled photons where each pair consists of a 785~nm wavelength signal photon and an 837~nm wavelength idler photon. For the purpose of QKD, each of the idler photons is measured aboard the satellite in either the computational or the diagonal basis with probability $1/2$. The signal photon is sent to the satellite's optical terminal using an optical interface. A subsystem inside the optical source also generates a synchronization beacon. Both the beacon and the signal photons are transmitted to the OGS through the satellite's optical terminal. 

Optical terminals on both the satellite and in the ground station help establish a space-to-ground free-space optical link. The terminals consist of optical telescopes and fine-pointing mechanisms for transmitting and collecting the signal photons, and synchronization and tracking beacons. Table~\ref{tab:SpooQy-OpticalLink} presents the parameters of the quantum source and the optical link. 

\newcommand*{\tabindent}{ \hspace{-1mm}}
\newlength{\thickarrayrulewidth}
\setlength{\thickarrayrulewidth}{2.1\arrayrulewidth}
\renewcommand{\arraystretch}{1.2}
\setlength{\tabcolsep}{8pt}
\begin{table}[t!]
  \centering
  \begin{tabular}{m{1.65cm}|m{1.5cm}|m{3.8cm}}
    \thickhline
    \tabindent\textbf{Parameter} & \tabindent\textbf{Value}  & \tabindent\textbf{Description} \\
    \hline
\tabindent\begin{tabular}[c]{@{}l@{}}Transmitter \\ aperture\end{tabular}   & \tabindent0.09 m          & \begin{tabular}[c]{@{}l@{}}\tabindent Realistic aperture size  \\ \tabindent for nanosatellite\end{tabular}                           \\ \hline
\tabindent\begin{tabular}[c]{@{}l@{}}Receiver \\ aperture\end{tabular}      & \tabindent0.6 m            & \tabindent\begin{tabular}[c]{@{}l@{}}Optimum aperture \end{tabular}       \\ \hline
\tabindent\begin{tabular}[c]{@{}l@{}}Beam \\ quality\end{tabular}              & \tabindent 1.6 M2           & \begin{tabular}[c]{@{}l@{}}\tabindent Fundamental limit is 1.4 \\ \tabindent due to diffraction\end{tabular}                          \\ \hline
\begin{tabular}[c]{@{}l@{}}\tabindent Pointing \\ \tabindent jitter\end{tabular}     & \tabindent 5 microrad              & \begin{tabular}[c]{@{}l@{}}\tabindent 1.2 microrad demonstrated \\ \tabindent on Micius satellite\end{tabular}                        \\ \hline
\tabindent Efficiency                                                            				   & \tabindent 50 \%             & \begin{tabular}[c]{@{}l@{}}\tabindent Estimated based on \\ \tabindent reflectivity and number \\ \tabindent of optical surfaces\end{tabular} \\ \hline
\tabindent \begin{tabular}[c]{@{}l@{}}Background \\ counts\end{tabular}       & \tabindent 1300 cps           & \begin{tabular}[c]{@{}l@{}}\tabindent Measured with respective \\ \tabindent setup in Singapore\end{tabular}                          \\ \hline
    \thickhline
  \end{tabular}
  \caption{Space-to-ground optical link parameters for CQT-Sat.}
  \label{tab:SpooQy-OpticalLink}
\end{table}%

\subsection{Emulating space-to-ground QKD using a tabletop setup}
\noindent
To emulate a space-to-ground QKD link we built the entanglement source and the detection apparatus representative of both the satellite and ground systems. The system parameters for this setup are listed in Table~\ref{tab:SpooQy-source-detector}.

\begin{figure}[b!]
    \centering
    \includegraphics[width=0.98\columnwidth]{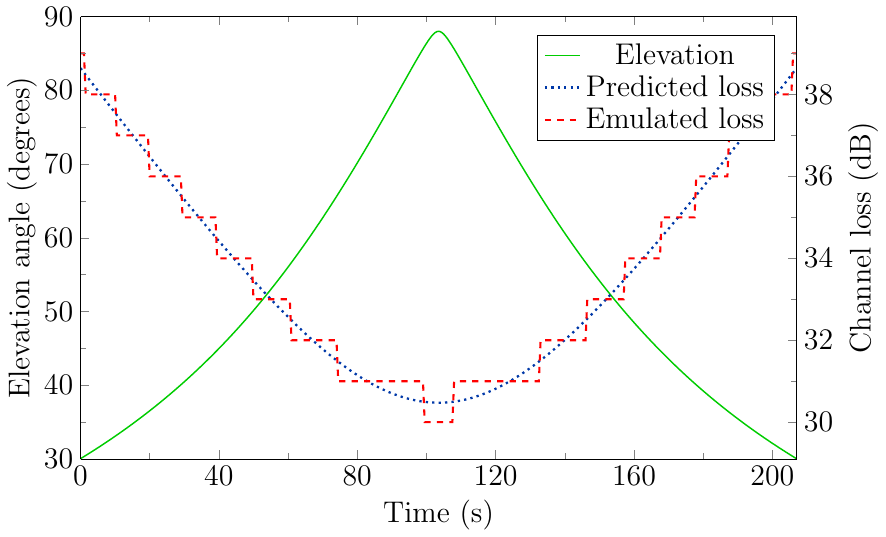}
    \caption{Example simulated satellite pass reaching $88^\circ$ elevation angle (green line). Experimental data is assembled such that its loss profile (red line) closely matches the theoretically predicted optical loss (blue line).} 
    \label{fig:spooqy-elev-loss}
\end{figure}

We consider a Sun-synchronous low Earth orbit with 500~km altitude above sea level. This orbit choice provides us with daily passes over the CQT ground station at a pre-specified time of the day~\cite{liao2017satellite}. We compute a time series of the satellite's angular elevation with respect to the OGS and the loss at that elevation for a pass using a simulation model that we have developed~\cite{vergoossen2020modelling,perumangatt2021realizing}. Regarding the simulation details, we compute the satellite range with respect to the OGS using the orbital simulation model and use parameters from Table~\ref{tab:SpooQy-source-detector} to compute the link loss at every point of the satellite pass. A satellite may pass over a ground station with a different maximum elevation. We simulate overpasses with a maximum elevation of $30^\circ$ to $90^\circ$. For example, in Figure~\ref{fig:spooqy-elev-loss} we show a pass with $88^\circ$ maximal elevation and associated loss that the optical link experiences. 

\begin{table}[]
\setlength{\tabcolsep}{14pt}
\renewcommand{\arraystretch}{1.2}
\begin{tabular}{l|l}
\hline
\textbf{System Parameter}                                          & \textbf{Value}   \\ \hline
Entangled pair production rate                                    & 20 Mcps \\ 
Source intrinsic QBER                                                	& 0.91 \% \\ 
Signal wavelength                                                       & 785 nm  \\ 
Idler wavelength                                                          & 837 nm  \\ 
Bandwidth                                                             	& 5 nm    \\ 
Detection efficiency                                                     & 25 \%    \\ 
Dark count rate per detector                                      	& 500 cps \\ 
Detector dead time                                                     	& 50 ns   \\ 
Detection jitter                                                            	& 320 ps  \\ 
Detector after-pulsing probability                               	& 5 \%    \\ \hline
\end{tabular}
\caption{\label{tab:SpooQy-source-detector} Source and detector parameters for CQT-Sat.}
\end{table}%

Using a variable attenuator we introduce different losses, and record detection timestamps for both signal and idler photons. Due to physical limitations, we only use a finite number of attenuator settings and stitch the experimental data together to emulate the predicted loss of the optical downlink. The blue and orange lines in Figure~\ref{fig:spooqy-elev-loss} illustrate the predicted and experimentally-emulated loss respectively at various segments of the satellite pass. 

This technique enables an investigation of satellite overpasses with different maximum elevations and to generate the associated detection timestamps both onboard the satellite and in the OGS. These timestamp sets are processed through the rest of the QKD protocol stack including finite key analysis to compute the secure key length achievable from each pass. A recent demonstration of a QKD system with a similar emulated satellite overpass was capable of establishing a 4.58-megabit secure key between two nodes~\cite{Roger2023_SA}.


\subsection{Key length of CQT-Sat for various LEO satellite passes}
\begin{figure}[ht]
    \centering
    \includegraphics[width=0.98\columnwidth]{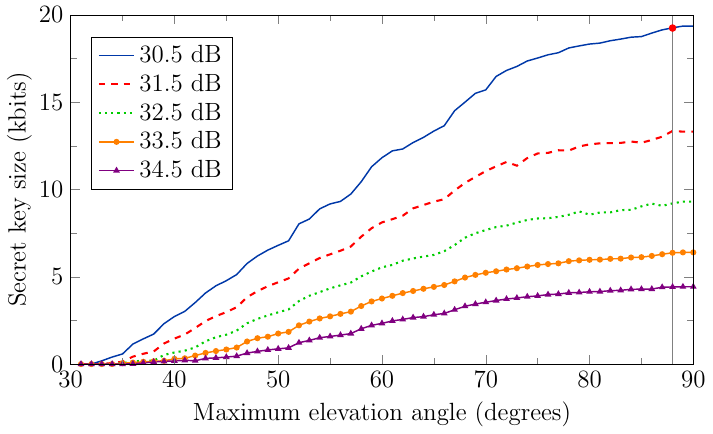}
    \caption{Secret key lengths achievable by CQT-Sat for passes of given maximal elevation. The blue curve gives results for the default simulation setting where the minimum loss at zenith is 30.5~dB. Other curves show the behaviour for optical links with additional 1 to 4~dB losses. All lines without markers are based on laboratory-based experimental data constructed to emulate pass-loss profiles. The results corresponding to the lines with markers contain loss profiles that are simulated numerically. The red marker on the blue line corresponds to the simulation depicted in Figure~\ref{fig:spooqy-elev-loss}, where the satellite pass has maximum elevation $88^\circ$.}
    \label{fig:finitekey}
\end{figure}

\noindent
Depending on geographical location and satellite orbit, a ground receiver might observe 2 to 6 satellite passes each day. An ideal satellite pass would transit directly over the ground station with maximal elevation of $90^\circ$ (zenith). At zenith, the satellite is closest to the OGS, and in clear weather this pass would exhibit the lowest transmission loss and longest link time.  However, such a pass is less likely than more ``glancing'' passes. For a given detector dark count rate, higher losses would result in a poorer signal-to-noise ratio and increase the QBER. Moreover, the pass duration and the number of photons successfully received from the satellite also decrease. Figure~\ref{fig:finitekey} shows how the secret key length changes with different satellite passes. Here we use the finite key analysis from~\cite{lim2021security} taking security parameter $10^{-10}$ where error correction efficiency is 1.18.

The analysis shows that below an elevation of $30^\circ$ no secret key is generated. This is acceptable for CQT-Sat which was designed to avoid operation at low elevation. The ground receiver in this case is sited at sea level in a tropical, urban environment and the optical channel below $30^\circ$ suffers more loss and light pollution due to the thicker atmospheric column. 

%


\section{Satellite-to-ground QKD using decoy-state BB84}
\label{sec:quarc_roks}
\noindent
The UK Quantum Research CubeSat (QUARC) project provides a design and architectural foundation for the Responsive Operations for Key Services (ROKS) mission in the National Space Innovation Programme (NSIP)~\cite{polnik2020scheduling, mazzarella2020quarc}. ROKS uses a continuation of the same 6U CubeSat platform as QUARC and will first implement decoy-state BB84 protocol in a downlink configuration for QKD service provision using a weak coherent pulse (WCP) source (Figure~\ref{fig:mission_configurations}). 

The satellite quantum modelling and analysis (SatQuMA) open-source software has been developed to estimate expected key generation performances for such satellite QKD missions~\cite{Satquma2021documentation}. SatQuMA models the efficient BB84 WCP two-decoy (three intensity) protocol and can optimise over the entire protocol parameter space and transmission segment time. It also incorporates recent results in finite-block composable secure key length estimation~\cite{Sidhu2020_npjQI, sidhu2021key,lim2021security}. SatQuMA can be applied to model the expected key generation performance for ROKS for a general satellite pass geometry in a Sun-synchronous orbit (SSO) at altitude $h$.


\subsection{System configuration}
\label{subsec:system_configuration}
\noindent
We use published empirical Micius mission measurements of the total optical loss of the SatQKD channel~\cite{Yin2020_N} to construct a representative total system link efficiency as a function of elevation angle during a satellite pass. To account for local horizon constraints around the OGS, we restrict quantum transmissions to elevations above $10^\circ$.

The link efficiency (loss) is highly dependent on the system parameters, OGS conditions, and orbits.  The nominal system parameters are summarized Table~\ref{tab:loss_error_parameters}, where the minimum total system loss at zenith is computed to be $34$ dB.  One can scale the minimum system loss at zenith to allow the comparison of differently performing SatQKD systems. Changes to the minimum system loss at zenith would then account for differences in the transmit and receive aperture sizes, pointing accuracy, atmospheric absorption, turbulence, receiver internal losses, and detector efficiencies. For the current simulations, we consider a nominal baseline value of 34~dB. SatQKD missions with differing performance can be modelled by linearly scaling the link efficiency vs elevation curve to account for different constant efficiency factors, such as a change in OGS receiver area.

To evaluate the sensitivity of the achievable secret key length to different errors, we categorise different contributions associated with sources and detectors in two key parameters. First, errors from dark counts and background light are combined together into a single extraneous count probability $p_\text{ec}$, here assumed to be constant and independent of elevation. In practice, it will depend strongly on the environment of the OGS and the light from celestial bodies.  Second, all other error terms, such as misalignment, source quality, and imperfect detection, are combined into an intrinsic quantum bit error rate $\text{QBER}_\text{I}$ independent of channel loss/elevation. This allows an efficient method to determine the sensitivity of the secret key length to different categories of errors, which helps identify targeted improvements for future SatQKD missions.

\setlength{\thickarrayrulewidth}{2.1\arrayrulewidth}
\renewcommand{\arraystretch}{1.2}
\setlength{\tabcolsep}{8pt}
\begin{table}[t!]
  \centering
  \begin{tabular}{m{2.3cm}|m{1.4cm}|m{3.3cm}}
    \thickhline
    \textbf{Parameter} 										& \textbf{Value} 	& \textbf{Description}\\
    \hline
    \begin{tabular}[c]{@{}l@{}}Intrinsic error\end{tabular}   				& 0.5\% 			& Source errors\\ 
    After-pulsing 												& 0.1\% 			& Probability of $p_\text{ap}$\\ 
    Extraneous count rate  										& $5\times 10^{-7}$ \vspace{2pt}	& Probability of counts from background light\\ 
    Source rate 												& 100 MHz		& Signal frequency\\ 
    Error correction \vspace{-8pt}											& $10^{-15}$ \vspace{2pt}		& Error-correction efficiency\\ 
    \begin{tabular}[c]{@{}l@{}}Security\end{tabular}					& $10^{-10}$ 		& Security parameter\\ 
    Altitude 													& $500$ km 		& Satellite orbit altitude\\ 
    System loss 												& $34$ dB 		& Loss at zenith\\ 
    \thickhline
  \end{tabular}
  \caption{Reference system parameters. We take published information of the Micius satellite and OGS system as representing an empirically derived set point for our finite key analysis. The total loss at zenith can be linearly scaled to model other systems with smaller OGSs or differing source rates.}
  \label{tab:loss_error_parameters}
\end{table}%
%


\subsection{Optimised finite key length}
\label{subsec:opt_finite_key}
\noindent
We model the efficient BB84 protocol, adopting the convention of key generation using signals encoded in the $\mathsf{X}$ basis and parameter estimation using signals in the $\mathsf{Z}$ basis, chosen with biased probabilities. For a two-decoy-state WCP BB84 protocol, one of three intensities $\mu_j$ for $j \in \{1,2,3\}$ are transmitted with probabilities $p_j$. An expression for the final finite key length, $\ell$, for this protocol is given in Ref.~\cite{Lim2014_PRA}.  The key is extracted from data for the whole pass as a single block without partitioning, the security proof of Ref.~\cite{Lim2014_PRA} makes no assumptions about the underlying statistics. This avoids having to combine small data blocks with similar statistics from different passes---thus, it is both quicker and avoids the need to track and store a combinatorially large number of link segments until each has attained a sufficiently large block size for asymptotic key extraction. 

The limited data sizes from restricted pass times results in key length corrections to account for finite statistics of the link. To improve the analysis, we use the tight multiplicative Chernoff bound~\cite{Yin2020_SR} and improve the estimate of error correction leakage $\lambda_\text{EC} \le \log \vert\mathcal{M}\vert$, where $\mathcal{M}$ characterises the set of error syndromes for reconciliation~\cite{Tomamichel2017_QIP} (see Ref.~\cite{Sidhu2020_npjQI} for more details).

\begin{figure}[t!]
    \centering
    \includegraphics[width=0.98\columnwidth]{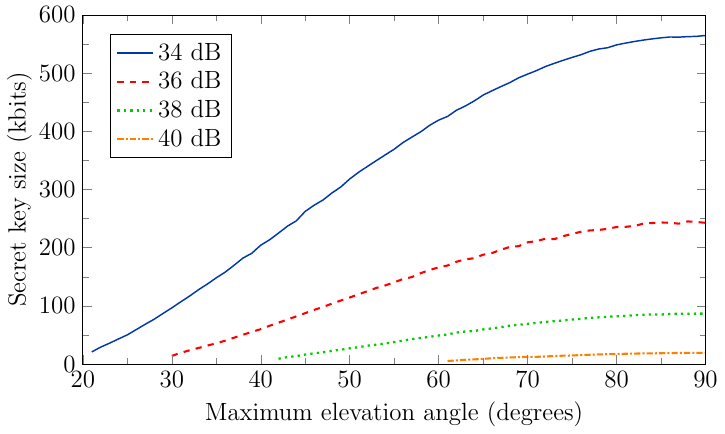}
    \caption{Estimated secret key length generated by QUARC mission for passes with different maximum elevation angles. In the default simulation the minimum total system loss is 34 dB. Other curves show behaviour for optical links with additional 2, 4, and 6~dB losses.}
    \label{fig:finitekey_QUARC}
\end{figure}%

For a defined SatQKD system, we optimise the finite key length $\ell$ by optimising over the protocol parameter space that includes the source intensities (with $\mu_3=0$) and their probabilities, and the basis encoding probability $p_\mathsf{X}$. We also optimise the portion of the pass data used for key generation.

The SatQuMA code is used to generate simulated measurement data.  The QBER and phase errors for the key bits are estimated using only the data from the complimentary basis. This is a classic sampling without replacement problem, which is usually solved in QKD using an approximation for the hypergeometric distribution \cite{fung2010}. Recently, however, an improved sampling bound has been proposed \cite{lim2021security}. This can be used to estimate the QBER and phase error. The formalism from Ref.~\cite{lim2021security} is used together with data generated from SatQuMA to determine the secret key length for different satellite passes, which we characterize through the maximum elevation angle.  The secret key is plotted in Figure~\ref{fig:finitekey_QUARC} as a function of the maximum elevation angle of a pass. 


\section{SatQKD using BBM92 in uplink or downlink configurations}

\noindent
The Quantum Encryption and Science Satellite (QEYSSat) mission~\cite{QEYSSat} is a Canadian initiative to develop and launch a microsatellite-hosted quantum receiver instrument into low-Earth orbit. The primary objective of the mission is to demonstrate QKD via quantum uplink from sources located at two or more ground stations. To support this, the QEYSSat instrument will possess a large front-end telescope for light collection, polarization discriminating optics, and single-photon avalanche diodes~\cite{Scott2020}. Support for a WCP downlink protocol is also being developed. As of writing, QEYSSat is in the late design/early construction phase and on schedule to launch in 2025.


\subsection{System configuration}

\begin{figure*}[t!]
    \centering
    \includegraphics[width=1\textwidth]{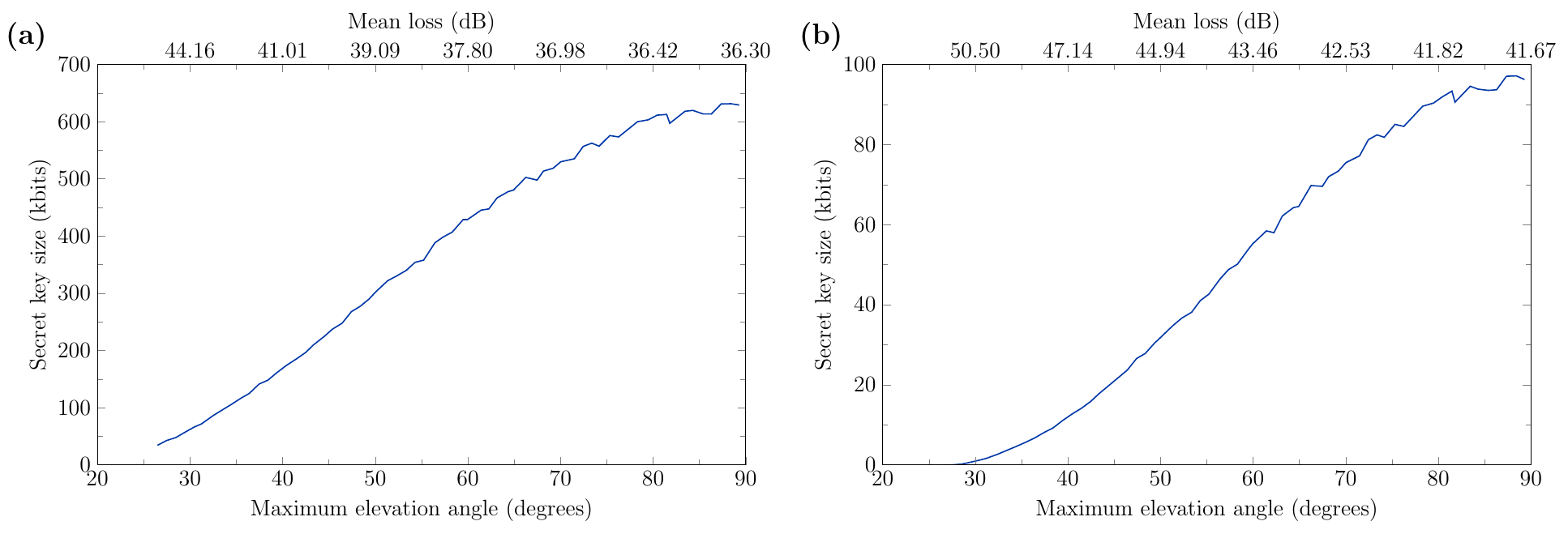}
    \caption{Expected performance under modelled conditions representative of a QEYSSat-style mission performing entanglement-based (BBM92) QKD in (a)~satellite-to-ground downlink configuration, and (b)~ground-to-satellite uplink configuration. Note that the QEYSSat mission baseline does not include entangled downlinks.}   
    \label{fig:mean and std of nets}
\end{figure*}%

\noindent
With QEYSSat's nominal configuration being an uplink, and the quantum sources located on the ground, lower secure key rates are expected when compared to a downlink with equivalent parameters. This is due to the steering effect of atmospheric turbulence on the beam at the beginning of its propagation in contrast to atmospheric steering at the end of propagation for a downlink channel. However, a satellite receiver affords considerably greater source flexibility. For this reason, two source types are baselined: WCP with decoy states in an unbiased BB84 protocol, and entangled photons (with one photon of each pair kept at the ground) in a BBM92 protocol. It is expected that other quantum source types---e.g., quantum dots (see, e.g., Ref.~\cite{Chaiwongkhot2022})---will also be employed during the experimental phase of the QEYSSat mission.

Commencement of the QEYSSat mission in 2018 was preceded by several theoretical and experimental investigations into the mission's feasibility, both as a whole~\cite{Jennewein2014} and with a focus on critical subsystems including pointing~\cite{Bourgoin2015,Pugh2017} and photon measurement~\cite{Anisimova2017,DSouza2021}. Of these, one early work~\cite{Bourgoin2013} numerically modelled the quantum optical link to establish the loss and fidelity of polarized-photon transmission under the assumptions of the expected orbital configuration and (generally conservative) atmospheric conditions. Multiple scenarios were considered, consisting of notional WCP or entangled-photon sources in both uplink and downlink configurations. Although some details of the in-development QEYSSat apparatus and conditions have been refined since, the values remain generally very similar.

In this section, we present the secret key generation performance of QEYSSat while executing the entanglement-based BBM92 protocol in both satellite-to-ground (downlink) and ground-to-satellite (uplink) quantum communication configuration modes. Although the QEYSSat instrument has ultimately been designed without entanglement downlink, we include its analysis here for two reasons: (1) for comparison with the uplink configuration that is planned, and (2) as an extension of the prior feasibility study performed in Ref.~\cite{Bourgoin2013}. We expect these updated results may influence future designs.


\subsection{Key length analysis}
\noindent
We calculate the secure key rate for the QEYSSat Mission using updated secure key length analysis~\cite{lim2021security}, which has improved performance with smaller raw key block size. Performance with smaller block size is important because it has implications on QKD feasibility under high-loss conditions and during low maximum elevation passes. This improved key length analysis enables higher key rates than prior analysis~\cite{Bourgoin2013}.

In this analysis, we set the error correction efficiency to 1.18 and the security parameter to $10^{-10}$, which is consistent with the values taken for the CQT-Sat and the QUARC/ROKS missions. Table~\ref{tab:QEYSSat-OpticalLink} summarizes the parameters that describe the quantum source and the optical link. The assumed satellite orbit (Sun-synchronous noon/midnight at \SI{600}{\km} altitude) was simulated for one year's duration of nighttime passes over a notional ground station located \SI{20}{\km} outside of Ottawa, Canada. Optical link conditions for each pass were modeled at ten-second intervals. The background light was determined from the Defence Meteorological Satellite Program's Operational Linescan System measurements~\cite{Elvidge1999, Cinzano2001, Bourgoin2013} and combined with an assumed half-moon at \SI{45}{\degree} (contributing via Earth reflection using its mean albedo) along with Earth's thermal (blackbody) radiation, taking into account the geometry of the optical field of view which changes over the pass of the satellite, and 1-nm-bandwidth spectral filtering. Detector dark counts of an additional \SI{20}{cps} were also included in the total noise detected.

\begin{table}
\begin{tabular}{l|l}
\hline
\textbf{Parameter} & \textbf{Value} \\
\hline
Orbital altitude & \SI{600}{\km} \\
Transmitter aperture & \SI{0.5}{\m} \\
Receiver aperture & \SI{0.3}{\m} \\
Pointing error & \SI{2}{microrad} \\
Optics losses & \SI{3}{\dB} \\
Quantum transmission wavelength & \SI{785}{\nm} \\
Detector loss & \SI{2.3}{\dB} \\
Spectral filtering bandwidth & \SI{1}{\nm} \\
Dark count rate per detector & \SI{20}{cps} \\
EPS pair production rate &  100 Mcps \\
Source intrinsic QBER & 1\% \\
Coincidence window & \SI{0.5}{\ns} \\
\hline
\end{tabular}
\caption{\label{tab:QEYSSat-OpticalLink}Ground-to-space optical link parameters for the model representing QEYSSat.}
\end{table}%

Here, we consider a transmitter and receiver diameter of \SI{50}{\cm} and \SI{30}{\cm} respectively as the baseline configuration. Optical losses were calculated from the contributions of numerically modeled diffraction given a central obstruction (secondary mirror), an assumed mean pointing error of \SI{2}{microrad}, atmospheric attenuation modeled by MODTRAN 5~\cite{berk2005modtran} for a ``rural'' profile with \SI{5}{\km} visibility, and Hufnagel--Valley model of atmospheric turbulence at sea-level. Photonic states were simulated in a 7-dimensional Fock-space (0 to 6 photons). Intrinsic reduction in quantum visibility was included via an operation equivalent to a small rotation.

Detector count rate statistics were calculated using assumed EPS pair production rate of \SI{100}{\mega pairs/s} via spontaneous parametric down-conversion (SPDC) pumped at $\epsilon=0.22$ (corresponding to a mean 0.1 pairs per pulse---see Ref.~\cite{Bourgoin2013}) with 98\% intrinsic entanglement visibility, and a \SI{0.5}{\ns} coincidence window. Such a source is challenging, but possible with current techniques on ground (for uplink), and can be reasonably foreseen as achievable with expected advances for space platforms (for downlink). Intervals, where the simulated measurement visibility was below 85\%, were filtered out (see Ref.~\cite{Erven2012}). In this analysis, we aggregate the remaining statistics at each pass and sorted these by the maximum elevation achieved by the satellite with respect to the ground station for that pass.

In Figure~\ref{fig:mean and std of nets}(a) and (b) we show the secret key generated for passes with different maximum elevation in the downlink and uplink configurations respectively for entanglement-based BBM92. We use the finite key analysis from Ref.~\cite{lim2021security}, with security parameter $10^{-10}$ and error correction efficiency 1.18 to compute the secure key lengths. Note that in comparison with the analysis done in Ref.~\cite{Bourgoin2013}, the secret key size is considerably greater---we expect this is largely a consequence of the faster source rate and assumed enhancements to intrinsic QBER and pointing accuracy, coupled with the highly nonlinear effect of finite-size statistical analysis.


\section{Mission comparisons}

\noindent
As the three SatQKD missions discussed here have different design specifications for the ground stations, satellites, and protocols implemented, a direct quantitative performance comparison of the missions is difficult. Despite this, we provide a qualitative discussion on the respective strengths and weaknesses of each mission. First, the uplink configuration employed by QEYSSat has the advantage that it relieves the source design from the strict size, weight, and power (SWaP) constraint imposed by a satellite. Moreover, it potentially allows QEYSSat to swap the entanglement source to a prepare-and-measure source at any point of the mission to perform a different SatQKD protocol and could benefit from abstract beam pointing. However, it has a disadvantage in that the optical link suffers larger environmental turbulence during the initial part of the optical path, generating higher beam wandering compared to a downlink configuration. 

The QUARC/ROKS mission uses a prepare-and-measure decoy state BB84 protocol where the source emits a periodic signal. This allows higher repetition rates to achieve larger raw key rates to counter the loss experienced in the satellite-to-ground optical link. In addition to a higher performance requirement for the detection and time-stamping circuits, the repetition rate is also constrained by the speed of quantum random number generation for the choice of quantum signal transmission. A similar attempt to counter the link loss by increasing the brightness of an entanglement source quickly hits a bottleneck because the SPDC-based source does not produce periodic signals, therefore due to the finite resolution of time-stamping devices and the limits imposed by detector jitters the source ends up producing too many multi-photon events per time-slot increasing the QBER to an unacceptable value. However, entanglement-based QKD implementations act as precursors to entanglement-sharing links, which are essential for future development toward a general-purpose quantum internet. 


\section{Outlook for global quantum communications}

\noindent
We compared three upcoming SatQKD missions in the preceding sections and show that an individual small satellite can satisfy finite-key requirements for SatQKD. For each of the three missions considered, we show this leads to non-zero finite keys generated for a single overpass. While an individual satellite in an appropriately chosen orbit can cover the Earth's surface each day, increased performance in the network will probably require constellations of these satellites~\cite{vergoossen2020modelling}. Aside from putting more satellites into space, it is important to consider how the performance of each individual satellite could be enhanced. We note from the preceding sections that LEO satellites operate at the edge of performance in terms of SatQKD that satisfy finite-key security. In this section, we report on specific challenges that, if overcome, can provide improved SatQKD performance over a wider range of operations. We also provide a long-term perspective on the demonstration of key milestones towards global quantum communications and applications beyond QKD.


\subsection{Outstanding challenges for SatQKD}
\label{subsec:challenges}

\noindent
Progress in finite-key security analyses presents an immediate and fundamental challenge to improving the achievable key rates. An improved finite-key analysis handles parameter estimation and post-processing tasks more efficiently. This would enable higher finite keys and successful distillation of non-zero finite keys at higher operating losses (e.g., lower elevation passes or at ground locations with worse atmospheres) without any hardware changes. Beyond improvements in finite-key analysis, there are specific challenges to hardware that would provide improved performances. 

A conventional research programme would revolve around improved transmitters and detectors. We propose that building a system that can operate effectively in daylight would be a major step. Practically, every SatQKD mission for the foreseeable future will operate during nighttime to avoid excess background light from the Sun. In order to operate during daylight, the spectral window of the transmitted light has to be sufficiently narrow for effective filtering, while the transmission system has to be built to avoid reflecting sunlight directly into the receiver. We note that adaptive optics (AO) for an optical ground receiver~\cite{acosta2021analysis,roberts2018first,wright2015adaptive}, to effectively couple light into a single mode fiber for direct transmission of the QKD signal to end-users located away from the ground receiver, will also become increasingly important. This has the added advantage that an AO system will act as a spatial filter, reducing the amount of stray light entering the quantum channel. To be useful, the AO system will need to be able to operate with high coupling efficiency, so that the overall system throughput is not compromised. A ground AO system can also improve uplink configurations~\cite{Pugh2020_AOT}. Research into transmission and detection systems that can penetrate cloud and fog would also be highly desirable.

Aside from the transmitter and detector aspects, we note that a major transmission loss contribution is the diffraction of the beam from the transmitter.
This loss could be mitigated using several different approaches. First is to operate the spacecraft at Very Low Earth Orbit (VLEO). This orbit has a nominal altitude below the International Space Station (approximately 400 km) and is often not considered due to satellites experiencing significant drag and re-entry into the Earth's atmosphere within a year. However, with space propulsion systems being developed for station keeping~\cite{scharlemann2011propulsion,Syarafana.2022}, this approach may become feasible and would afford significantly lower losses owing to shorter optical links. In designing a VLEO system, factors such as micro-buffeting from the residual atmosphere, degradation from atomic oxygen, and the shorter overhead time of the satellite remain open challenges.

\begin{table}[t!]
\begin{tabular}{l|l}
\hline
\textbf{Parameter} & \textbf{Value} \\
\hline
Beam waist to aperture ratio & 0.89 \\
Satellite transmitter aperture & \SI{0.1}{\m} \\
Pointing error & \SI{2.5}{microrad}\\
Ground receiver aperture & \SI{0.6}{\m}\\
Wavelength & \SI{785}{\nm}\\
\hline
\end{tabular}
\caption{\label{tab:sim_table} Space-to-ground optical link parameters used for modeling higher altitude orbits.}
\end{table}%

Second is the use of enhanced transmit/receive apertures. The use of larger apertures has been the primary route to minimise link loss, with a doubling in aperture sizes providing 6~dB improvements~\cite{Sidhu2020_npjQI}. However, aperture sizes are restricted for small satellites. Recent efforts rely on deployable and active optics~\cite{Schwartz2016, Corbacho2020}. 

Finally, diffraction losses can also be mitigated by developing larger and more capable satellites at very high altitudes. The advantage is that such satellites can be equipped with large transmit apertures while increasing the ground coverage area as well as improving access time for a ground receiver. The drawback is a dramatic increase in diffraction loss that must be compensated by enlarging the transmit/receive aperture and improving pointing accuracy.

\begin{figure}[t!]
    \centering
    \includegraphics[width=0.98\columnwidth]{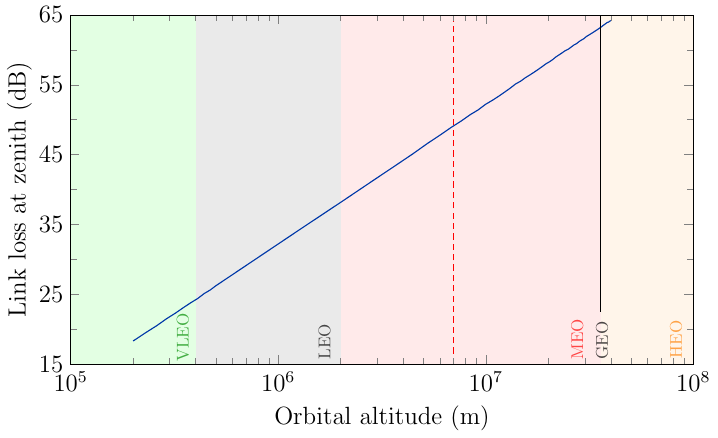}
    \caption{Link loss as a function of orbital altitude using parameters from Table~\ref{tab:sim_table}. Green shaded region indicates the altitude range for VLEO orbits, gray for LEO orbits, red for MEO orbits, orange for HEO, and the solid black line for a GEO orbit. The red dashed line corresponds to a representative MEO altitude of $7 \times 10^3$ km that we consider later. Link loss at zenith rapidly increases with increasing orbital altitudes.}
    \label{fig:altitude_vs_linkloss}
\end{figure}

\begin{figure*}[t!]
    \centering
    \includegraphics[width=0.91\textwidth]{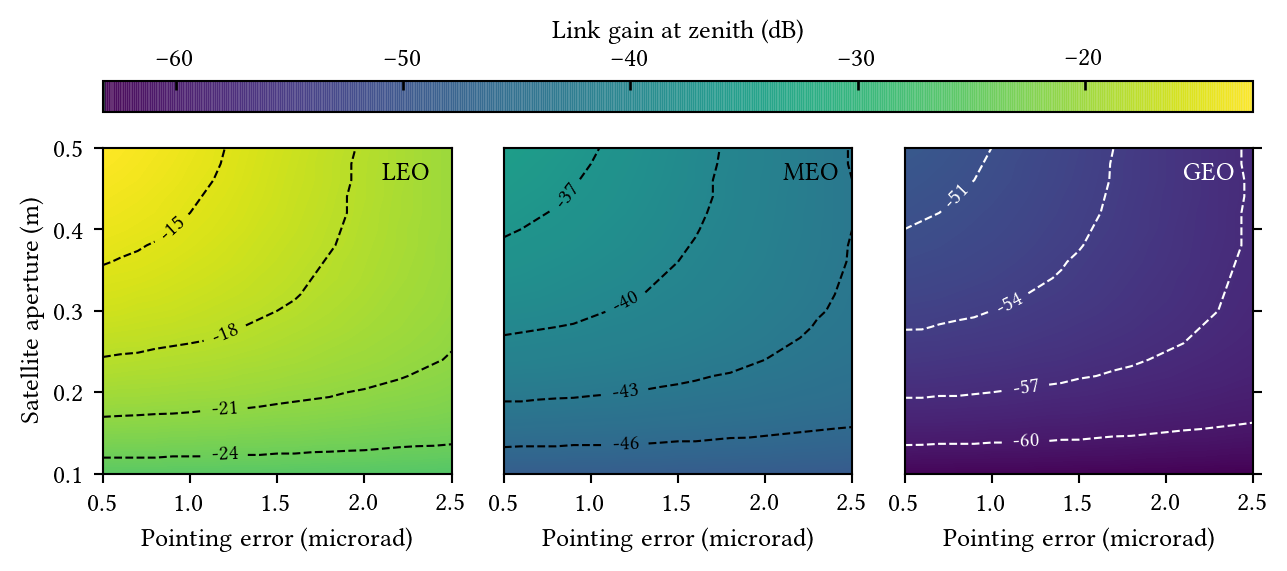}
    \caption{Trade-off between telescope aperture onboard the satellite and pointing error with respect to the link gain at zenith. (a) The trade-off for a representative low Earth orbit (LEO) at an altitude of 500 km. (b) The trade-off for a representative medium Earth orbit (MEO) at an altitude of 7000 km. (c) The trade-off for a geostationary orbit (GEO).}
    \label{fig:challenges}
\end{figure*}

We have modelled the performance of SatQKD systems for varying orbital altitudes, by imposing similar capabilities on the satellites for LEO, Medium Earth Orbit (MEO)~\cite{PhysRevA.93.010301} and Geostationary Orbit (GEO)~\cite{dirks2021geoqkd}.
Under a downlink configuration using the receiver (ground-based) and sender (onboard-satellite) telescope and beam parameters as shown in Table~\ref{tab:sim_table}, we study the space-to-ground optical link loss (Figure~\ref{fig:altitude_vs_linkloss}) for higher altitude orbits. We see that the optical link loss increases rapidly with the increase in altitude as expected from the beam expansion. One approach to counter increased losses at higher orbits and ensure successful operation of SatQKD is to use ultra-bright sources capable of operating at GHz-bandwidth repetition rates~\cite{ecker2021strategies}. SatQKD missions typically operate with a source rate in the order of $10^8$ Hz. Increasing the source rate to the GHz-range and beyond requires low timing jitters that are on the sub-ns scale. This is possible with superconducting nanowire single-photon detectors (SNSPDs)~\cite{Holzman2019_AQT} at the expense of greater SWaP due to the requirement of cryogenic cooling. An alternative approach to compensating the increased loss at higher orbits is to increase parameters such as the sending or receiving telescope apertures and pointing accuracy. In Figure~\ref{fig:challenges} we show a trade-off heat-map where the satellite's transmitter telescopes aperture and pointing accuracy are varied to show how it affects the transmission for a satellite in LEO, MEO, and GEO. These trade-off calculations show that for orbits higher than LEO it is not sufficient to only change the satellite's parameter to achieve the transmission gain necessary for successful SatQKD. For higher altitudes, one would also need to improve other parameters, such as the ground telescope's aperture, pointing accuracy, and detector performance.


\begin{figure}[b!]
    \centering
    \includegraphics[width=0.99\columnwidth]{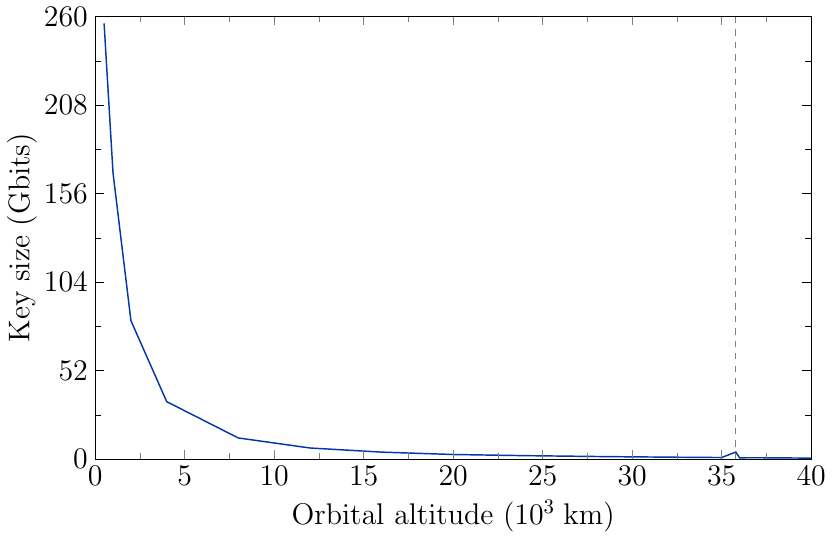}
    \caption{Raw key length accumulated over a year as a function of orbital altitude. The gray dashed vertical line indicates a GEO altitude, where the satellite remains stationary at the zenith leading to a higher raw key and lower QBER compared to other nearby orbits where the satellite has varying angular elevation with respect to the OGS during an overpass. To model this, we assume a satellite telescope with aperture \SI{0.5}{m}, OGS telescope with aperture \SI{1.8}{m} and the entangled pair production rate 100 Mcps.} 
    \label{fig:accumulation_1_year}
\end{figure}

Due to increased losses at higher altitude orbits, obtaining a secure key from a single pass over a ground station may not be possible in these cases. It is possible to accumulate the raw key bits over multiple passes, increasing the block size to reduce finite block size effects sufficiently to achieve a secure key. In Figure~\ref{fig:accumulation_1_year}, we show block sizes and associated QBERs for the raw key bits accumulated over one year for entanglement-based SatQKD operated at various orbital altitudes for a single link. The drawback of key aggregation is that large amounts of data will have to be stored for a long time onboard the satellite which might introduce vulnerabilities due to storage security. In Figure~\ref{fig:accumulation_1_year}, we discard passes that yield QBER higher than $11\%$, which generates a key with minimal information known to an eavesdropper after post-processing~\cite{Lutkenhaus1999estimates}. However, given the limited number of bits acquired at each pass it might not be feasible to determine the QBER reliably by exchanging a subset of these bits between the satellite and the OGS.  

Missions that choose to implement larger operating apertures to counter larger losses from higher-altitude orbits should also consider the increased costs associated with the optics and mass of the satellite. Specifically, the estimated cost variation for larger telescopes is $T_x^{1.7}$~\cite{Stahl2019Multivariable} largely due to bulk optics and increased mass, making them considerably more expensive than smaller telescopes. Further, space-based telescopes are estimated to be 30 times more expensive than ground telescopes.

%


\subsection{Long-term perspectives}
\label{subsec:long_term_perspectives}

\noindent
In this section, we provide a perspective on medium to long-term challenges and milestones that present a blueprint for enabling additional capabilities for the quantum internet. These milestones relate to the implementation of different QKD protocols to extend the use cases of quantum communications, and the development of improved hardware that could enable the demonstration of distributed quantum technologies beyond communications.

To extend the range of quantum communications, the horizon of efforts in developing SatQKD systems is likely to involve the improvement of instrument components such as the sources, detectors, classical communication systems, and optical systems. This would directly generate improved key rates and would enable the implementation of a number of additional QKD protocols. 

First, the development of quantum memories will provide synchronization of probabilistic events to enable the implementation of memory-assisted (MA)-QKD protocols. Recent theoretical studies have shown that MA-QKD protocols can yield higher key rates over global distances and provides improved robustness against atmospheric weather and multiple-excitation effects~\cite{Nicolo2017_IOP,Gundogan2021npjQI,Wallnofer2021}. The advantage is that MA-QKD is less demanding on the performances of quantum memories than those required for probabilistic quantum repeaters. Demonstration of MA-QKD protocols can herald a major progress towards improved rate-versus-distance performance of SatQKD. Beyond communications, satellite-based quantum memories can enable distributed quantum sensing and imaging~\cite{Sidhu2017_PRA, Sidhu2018_arxiv, Sidhu2021_PRX}. However, any distributed applications making use of quantum memories will require active tracking and compensation of Doppler shifts that arise from the rapid relative motion of satellites, which provides a further challenge. For example, the typical speed of a LEO satellite is 7800 ms$^{-1}$, which has a maximum fractional Doppler shift (generally, elevation dependent) of $\beta = v/c =  2.6\times10^{-5}$. Compensating for this shift is important to enable the signal to efficiently couple with a narrow line width of the quantum memory. Compensation for Doppler shifts may also require inter-conversion between flying and static quantum systems.

Second, continuous-variable (CV) protocols operate with conventional telecommunication devices and homodyne measurements that can be implemented at room temperature. This improves integration with existing ground-based networks and circumvents the need for bulky systems to cryogenically cool single-photon detectors which may be necessary to support the discrete variable (DV) protocols, including BB84 and BBM92, that have been our focus. Despite these advantages, CV-QKD provides limited key rates at high-loss, large distances that are typical in SatQKD, which explains their limited use in proposed SatQKD. There are studies that explore the feasibility of CV QKD over large distances~\cite{He2019, Dequal2021_npjQI}. In addition, CV-QKD does not have the same maturity of proof techniques in the finite key regime. This in particular leaves an open question about the security of finite keys and how keys can be more efficiently distilled for overpass data. Initial studies on this provide an optimistic outlook~\cite{Dequal2021_npjQI}, but further studies are required to establish the same maturity and rigour that DV QKD protocols have. Particular challenges include determining the optimal approach to partition overpass data and finding analytic finite-key methods that do not rely on numerical approximations.

Third, if the polarization degree of freedom of a photon is used to encode qubit in a SatQKD link (as is common), it becomes crucial that both the sender and the receiver have their reference frame well aligned, as misalignment results in quantum bit errors in the output~\cite{bartlett2007reference}. The relative motion between the satellite and the ground station may introduce additional challenges in acquiring and maintaining a common reference frame during quantum communication. Several reference frame independent quantum communication protocols~\cite{laing2010reference,wang2015reference,tannous2019demonstration, Jin2018_PRA, Jin2019_OE, Tannous2023_arxiv} have also been proposed to account for this. 

Finally, time-bin encoding offers another way of encoding key bits in DV-QKD~\cite{tittel2000}. In principle, time-bin encoding can allow more than one bit to be encoded on each photon~\cite{islam2017, frameqkd2016}.  The motivation for this approach is that increasing data rates would normally require an increase in the source repetition rate, but this eventually reaches a practical limit due to detector dead time.  Higher-dimensional encodings could circumvent this issue by allowing multiple bits to be encoded on each photon.  Currently, such systems have only been demonstrated in laboratory settings~\cite{islam2017}. Considerable work remains to develop setups that could be deployed in the challenging conditions of a satellite.  Further, many of the QKD systems investigated to date use fiber, not free space. Only recently have there been investigations into the effects of errors due to free-space transmission in such setups~\cite{castillo2020}.  

Phase-randomised weak-coherent-pulses have become the most well-studied and implemented information carriers in satellite-based missions due to their maturity and ease of implementation~\cite{Sidhu2021advances}. However, recent progress in quantum source development provides access to alternative QKD sources. For example, true single photon source (SPS) based on Nitrogen-vacancy centers~\cite{alleaume2004experimental} and quantum dots~\cite{takemoto2015quantum} are being developed and may become suitable for small SatQKD applications in the near future. The use of SPSs would provide enhanced security given their inherent immunity to photon number splitting attacks, and would also provide advantages in general purpose quantum communications such as quantum repeaters~\cite{sangouard2012single}, optical quantum memory~\cite{Lvovsky2009_NP}, and on-demand entanglement generation~\cite{muller2014demand}.

For applications beyond QKD, small satellites could benefit by using multiple, independently steerable telescopes to distribute entanglement to multiple OGSs. This will help minimise latency in distributed quantum technologies when multiple ground stations are in view. Although steering multiple telescopes on small satellites increases the mechanical complexity and mass of the instrument, possible disturbance to the alignment of optical systems could be mitigated with twin tethered nanosatellites. This naturally raises the possibility of formation flying of small satellite clusters that can extend the range of applications. This would be particularly important for distributed quantum technologies.

A LEO satellite is inherently limited in the geographical area it can cover at any one time. A satellite-based global quantum network will therefore need a constellation of satellites~\cite{vergoossen2020modelling} that may involve small-satellite-based quantum communication between satellites~\cite{naughton2019design} and other high-altitude flying platforms~\cite{islam2017approaches}. Along with the DV SatQKD systems described in this work, there are studies~\cite{dequal2021feasibility} investigating the feasibility of continuous variable satellite-based QKD systems. 


\section{Conclusion}

\noindent 
There is growing interest in deploying satellites to enable a global QKD network. To ensure this goal remains feasible and to guide experimental and engineering efforts, it is crucial to understand how SatQKD can yield efficient secret key generation under finite transmission times and high loss regimes. Previous works have shown that secret key generation with SatQKD is possible using finite-key analyses. Recent advancements in the treatment of finite key effects have improved the efficiency of key extraction, which greatly decreases the requirements on the minimum raw key length necessary for key extraction.

We use these latest finite key bounds in the performance analyses of three different LEO-based satellite mission concepts; the 12U CQT-Sat mission implementing an entanglement-based BBM92 downlink, the 6U QUARC project implementing a WCP decoy-state BB84 in downlink configuration, and the QEYSSat mission implementing both the decoy BB84 and BBM92 protocols in an uplink configuration, in addition to a decoy BB84 state downlink. All three SatQKD missions achieve good secure key yields on the order of kilobits from a single pass over a ground receiver, even for the missions based on resource-constrained and aperture-limited CubeSats. This provides reassurance that planned SatQKD missions are on course to achieve important milestones that can lead to an effective global QKD network.

The long-term vision of a satellite-based global quantum network remains a principal motivation behind SatQKD. Therefore, developing the infrastructure for a global QKD network sets the stage for future theoretical, experimental, and engineering milestones. We list these milestones together with outstanding challenges in the field and discuss potential routes to overcome them. Prominent challenges discussed include the daylight operation of SatQKD, the cooperation of multiple OGSs with a constellation of satellites to improve the reliability of general applications beyond QKD services, and implementing SatQKD from different altitudes to enable longer-range communications and inter-satellite links. We extend our discussion by modelling the performance of SatQKD systems with varying orbital altitudes and quantify system design trade-offs to offset the increased link losses at higher altitudes. While our calculations demonstrate that all three LEO SatQKD missions considered here have the ability to yield secure finite keys, it is clear that implementing SatQKD from higher altitudes require overcoming numerous hardware challenges and further improving security analyses simultaneously.

For applications beyond QKD, the most demanding technological challenge is to implement general purpose quantum communications that has applications in distributed quantum technologies, such as quantum computing, error correction, and quantum sensing. This will require a constellation of satellites, each synchronised and equipped with entanglement sources and quantum memories to dynamically create multi-link connections between any two points on Earth. Our discussion on the short and long-term perspectives of satellite based quantum communications help build a blueprint for enabling the global quantum internet.

\acknowledgments

\noindent
J.S.S, T.B., and D.K.L.O. acknowledge the support of the UKNQTP and the Quantum Technology Hub in Quantum Communications (EPSRC Grant EP/T001011/1). B.L.H. and T.J. acknowledge support from the Canadian Space Agency, and thank J.-P. Bourgoin for discussions. T.I. and A.L. acknowledge support from the Research Centres of Excellence programme supported by the National Research Foundation (NRF) Singapore and the Ministry of Education, Singapore. This work was supported by the EPSRC International Network in Space Quantum Technologies (grant ref: EP/W027011/1).

\bibliographystyle{ieeetr}

\begin{thebibliography}{10}

\bibitem{Donkor2004distributed}
A.~Yimsiriwattana and S.~J.~L. Jr., ``Distributed quantum computing: a
  distributed shor algorithm,'' in {\em Quantum Information and Computation II}
  (E.~Donkor, A.~R. Pirich, and H.~E. Brandt, eds.), vol.~5436, pp.~360 -- 372,
  SPIE, 2004.

\bibitem{Meter2016the}
R.~Van~Meter and S.~J. Devitt, ``The path to scalable distributed quantum
  computing,'' {\em Computer}, vol.~49, no.~9, pp.~31--42, 2016.

\bibitem{Liorni2021}
C.~Liorni, H.~Kampermann, and D.~Bru{\ss}, ``Quantum repeaters in space,'' {\em
  New J. Phys.}, vol.~23, p.~053021, may 2021.

\bibitem{Wallnofer2021}
J.~Walln{\"o}fer, F.~Hahn, M.~G{\"u}ndo{\u{g}}an, J.~S. Sidhu, F.~Kr{\"u}ger,
  N.~Walk, J.~Eisert, and J.~Wolters, ``Simulating quantum repeater strategies
  for multiple satellites,'' {\em Communications Physics}, vol.~5, no.~1,
  p.~169, 2022.

\bibitem{Sidhu2020_AVS}
J.~S. Sidhu and P.~Kok, ``Geometric perspective on quantum parameter
  estimation,'' {\em AVS Quantum Science}, vol.~2, p.~014701, February 2020.

\bibitem{Gundogan2021npjQI}
M.~G{\" u}ndo{\u g}an, J.~S. Sidhu, V.~Henderson, L.~Mazzarella, J.~Wolters,
  D.~K. Oi, and M.~Krutzik, ``Proposal for space-borne quantum memories for
  global quantum networking,'' {\em npj Quantum Information}, vol.~7, p.~128,
  August 2021.

\bibitem{gundogan2021topical}
M.~G{\" u}ndo{\u g}an, T.~Jennewein, F.~K. Asadi, E.~D. Ros, E.~Sa{\u g}lamy{\"
  u}rek, D.~Oblak, T.~Vogl, D.~Riel{\" a}nder, J.~Sidhu, S.~Grandi,
  L.~Mazzarella, J.~Wallnöfer, P.~Ledingham, L.~LeBlanc, M.~Mazzera,
  M.~Mohageg, J.~Wolters, A.~Ling, M.~Atat{\" u}re, H.~de~Riedmatten, D.~Oi,
  C.~Simon, and M.~Krutzik, ``Topical white paper: A case for quantum memories
  in space,'' {\em arXiv:2111.09595}, November 2021.

\bibitem{Gundogan2023time}
M.~G{\" u}ndo{\u g}an, J.~S. Sidhu, M.~Krutzik, and D.~K.~L. Oi, ``Time-delayed
  single quantum repeater node for global quantum communications with a single
  satellite,'' {\em arXiv e-prints}, p.~2303.04174v2, September 2023.

\bibitem{Sidhu2021advances}
J.~S. Sidhu, S.~K. Joshi, M.~Gündoğan, T.~Brougham, D.~Lowndes,
  L.~Mazzarella, M.~Krutzik, S.~Mohapatra, D.~Dequal, G.~Vallone, P.~Villoresi,
  A.~Ling, T.~Jennewein, M.~Mohageg, J.~G. Rarity, I.~Fuentes, S.~Pirandola,
  and D.~K.~L. Oi, ``Advances in space quantum communications,'' {\em IET
  Quantum Communication}, vol.~2, no.~4, pp.~182--217, 2021.

\bibitem{belenchia2021quantum}
A.~Belenchia, M.~Carlesso, {\" O}.~Bayraktar, D.~Dequal, I.~Derkach,
  G.~Gasbarri, W.~Herr, Y.~L. Li, M.~Rademacher, J.~Sidhu, D.~K. Oi, S.~T.
  Seidel, R.~Kaltenbaek, C.~Marquardt, H.~Ulbricht, V.~C. Usenko,
  L.~W{\"o}rner, A.~Xuereb, M.~Paternostro, and A.~Bassi, ``Quantum physics in
  space,'' {\em Physics Reports}, vol.~951, pp.~1--70, 2022.

\bibitem{wehner2018quantum}
S.~Wehner, D.~Elkouss, and R.~Hanson, ``Quantum internet: A vision for the road
  ahead,'' {\em Science}, vol.~362, no.~6412, p.~eaam9288, 2018.

\bibitem{jianwei2018progress}
P.~Jianwei, ``{Progress of the quantum experiment science satellite (QUESS)
  Micius project},'' {\em Chin. J. Space Science}, vol.~38, no.~5,
  pp.~604--609, 2018.

\bibitem{villar2020entanglement}
A.~Villar, A.~Lohrmann, X.~Bai, T.~Vergoossen, R.~Bedington, C.~Perumangatt,
  H.~Y. Lim, T.~Islam, A.~Reezwana, Z.~Tang, {\em et~al.}, ``Entanglement
  demonstration on board a nano-satellite,'' {\em Optica}, vol.~7, no.~7,
  pp.~734--737, 2020.

\bibitem{Sidhu2023finite}
J.~S. Sidhu, T.~Brougham, D.~McArthur, R.~G. Pousa, and D.~K.~L. Oi, ``Finite
  key performance of satellite quantum key distribution under practical
  constraints,'' {\em Communications Physics}, vol.~6, p.~210, August 2023.

\bibitem{Sidhu2023satellite}
J.~S. Sidhu, T.~Brougham, D.~McArthur, R.~G. Pousa, and D.~K.~L. Oi,
  ``{Satellite quantum key distribution performance analysis and optimization
  with finite key size constraints},'' in {\em Quantum Computing,
  Communication, and Simulation III} (P.~R. Hemmer and A.~L. Migdall, eds.),
  vol.~12446, p.~124460M, International Society for Optics and Photonics, SPIE,
  2023.

\bibitem{Lim2014_PRA}
C.~C.~W. Lim, M.~Curty, N.~Walenta, F.~Xu, and H.~Zbinden, ``Concise security
  bounds for practical decoy-state quantum key distribution,'' {\em Phys. Rev.
  A}, vol.~89, p.~022307, February 2014.

\bibitem{Tomamichel2017_QIP}
M.~Tomamichel, J.~Martinez-Mateo, C.~Pacher, and D.~Elkouss, ``Fundamental
  finite key limits for one-way information reconciliation in quantum key
  distribution,'' {\em Quant. Inf. Proc.}, vol.~16, p.~280, October 2017.

\bibitem{Zhang2017_PRA}
Z.~Zhang, Q.~Zhao, M.~Razavi, and X.~Ma, ``Improved key-rate bounds for
  practical decoy-state quantum-key-distribution systems,'' {\em Phys. Rev. A},
  vol.~95, p.~012333, January 2017.

\bibitem{Yin2020_SR}
H.-L. Yin, M.-G. Zhou, J.~Gu, Y.-M. Xie, Y.-S. Lu, and Z.-B. Chen, ``Tight
  security bounds for decoy-state quantum key distribution,'' {\em Sci. Rep.},
  vol.~10, p.~14312, August 2020.

\bibitem{lim2021security}
C.~C.-W. Lim, F.~Xu, J.-W. Pan, and A.~Ekert, ``Security analysis of quantum
  key distribution with small block length and its application to quantum space
  communications,'' {\em Physical Review Letters}, vol.~126, no.~10, p.~100501,
  2021.

\bibitem{Bourgoin2015b}
J.-P. Bourgoin, N.~Gigov, B.~L. Higgins, Z.~Yan, E.~Meyer-Scott, A.~K.
  Khandani, N.~L\"utkenhaus, and T.~Jennewein, ``Experimental quantum key
  distribution with simulated ground-to-satellite photon losses and processing
  limitations,'' {\em Phys.\ Rev.\ A}, vol.~92, p.~052339, 2015.

\bibitem{anwar2021entangled}
A.~Anwar, C.~Perumangatt, F.~Steinlechner, T.~Jennewein, and A.~Ling,
  ``Entangled photon-pair sources based on three-wave mixing in bulk
  crystals,'' {\em Review of Scientific Instruments}, vol.~92, no.~4,
  p.~041101, 2021.

\bibitem{perumangatt2021realizing}
C.~Perumangatt, T.~Vergoossen, A.~Lohrmann, S.~Sivasankaran, A.~Reezwana,
  A.~Anwar, S.~Sachidananda, T.~Islam, and A.~Ling, ``Realizing quantum nodes
  in space for cost-effective, global quantum communication: in-orbit results
  and next steps,'' in {\em Quantum Computing, Communication, and Simulation},
  vol.~11699, p.~1169904, SPIE, 2021.

\bibitem{liao2017satellite}
S.-K. Liao, W.-Q. Cai, W.-Y. Liu, L.~Zhang, Y.~Li, J.-G. Ren, J.~Yin, Q.~Shen,
  Y.~Cao, Z.-P. Li, {\em et~al.}, ``Satellite-to-ground quantum key
  distribution,'' {\em Nature}, vol.~549, no.~7670, pp.~43--47, 2017.

\bibitem{vergoossen2020modelling}
T.~Vergoossen, S.~Loarte, R.~Bedington, H.~Kuiper, and A.~Ling, ``Modelling of
  satellite constellations for trusted node qkd networks,'' {\em Acta
  Astronautica}, vol.~173, pp.~164--171, 2020.

\bibitem{Roger2023_SA}
T.~Roger, R.~Singh, C.~Perumangatt, D.~G. Marangon, M.~Sanzaro, P.~R. Smith,
  R.~I. Woodward, and A.~J. Shields, ``Real-time gigahertz free-space quantum
  key distribution within an emulated satellite overpass,'' {\em Science
  Advances}, vol.~9, no.~48, p.~eadj5873, 2023.

\bibitem{polnik2020scheduling}
M.~Polnik, L.~Mazzarella, M.~Di~Carlo, D.~K. Oi, A.~Riccardi, and
  A.~Arulselvan, ``Scheduling of space to ground quantum key distribution,''
  {\em EPJ Quantum Technology}, vol.~7, no.~1, p.~3, 2020.

\bibitem{mazzarella2020quarc}
L.~Mazzarella, C.~Lowe, D.~Lowndes, S.~K. Joshi, S.~Greenland, D.~McNeil,
  C.~Mercury, M.~Macdonald, J.~Rarity, and D.~K.~L. Oi, ``Quarc: Quantum
  research cubesat—a constellation for quantum communication,'' {\em
  Cryptography}, vol.~4, no.~1, p.~7, 2020.

\bibitem{Satquma2021documentation}
J.~S. Sidhu, T.~Brougham, D.~McArthur, R.~G. Pousa, and D.~K.~L. Oi,
  ``Satellite quantum modelling \& analysis software version 1.1:
  Documentation,'' {\em arXiv: 2109.01686}, January 2022.

\bibitem{Sidhu2020_npjQI}
J.~S. Sidhu, T.~Brougham, D.~McArthur, R.~G. Pousa, and D.~K.~L. Oi, ``Finite
  key effects in satellite quantum key distribution,'' {\em npj Quant. Inf.},
  vol.~8, no.~1, p.~18, 2022.

\bibitem{sidhu2021key}
J.~S. Sidhu, T.~Brougham, D.~McArthur, R.~G. Pousa, and D.~K.~L. Oi, ``{Key
  generation analysis for satellite quantum key distribution},'' in {\em
  Quantum Technology: Driving Commercialisation of an Enabling Science II}
  (M.~J. Padgett, K.~Bongs, A.~Fedrizzi, and A.~Politi, eds.), vol.~11881,
  pp.~1 -- 8, International Society for Optics and Photonics, SPIE, 2021.

\bibitem{Yin2020_N}
J.~Yin, Y.-H. Li, S.-K. Liao, M.~Yang, Y.~Cao, L.~Zhang, J.-G. Ren, W.-Q. Cai,
  W.-Y. Liu, S.-L. Li, {\em et~al.}, ``Entanglement-based secure quantum
  cryptography over 1,120 kilometres,'' {\em Nature}, vol.~582, no.~7813,
  pp.~501--505, 2020.

\bibitem{fung2010}
C.-H.~F. Fung, X.~Ma, and H.~F. Chau, ``Practical issues in
  quantum-key-distribution postprocessing,'' {\em Phys. Rev. A}, vol.~81,
  p.~012318, Jan 2010.

\bibitem{QEYSSat}
``{Quantum EncrYption and Science Satellite (QEYSSat)}.''
  \url{http://qeyssat.ca/}, 2022.

\bibitem{Scott2020}
A.~Scott, T.~Jennewein, J.~Cain, I.~D'Souza, B.~Higgins, D.~Hudson, H.~Podmore,
  and W.~Soh, ``The {QEYSS}at mission: On-orbit demonstration of secure optical
  communications network technologies,'' {\em Proc. SPIE}, vol.~11532,
  p.~115320H, 2020.

\bibitem{Chaiwongkhot2022}
P.~Chaiwongkhot, S.~Hosseini, A.~Ahmadi, B.~L. Higgins, D.~Dalacu, P.~Poole,
  R.~L. Williams, M.~E. Reimer, and T.~Jennewein, ``Enhancing secure key rates
  of satellite {QKD} using a quantum dot single-photon source,'' {\em
  arXiv:2009.11818}, 2020.

\bibitem{Jennewein2014}
T.~Jennewein, J.-P. Bourgoin, B.~Higgins, C.~Holloway, E.~Meyer-Scott,
  C.~Erven, B.~Heim, Z.~Yan, H.~H\"ubel, G.~Weihs, E.~Choi, I.~D'Souza,
  D.~Hudson, and R.~Laflamme, ``{QEYSSAT}: a mission proposal for a quantum
  receiver in space,'' {\em Proc. SPIE}, vol.~8997, p.~89970A, 2014.

\bibitem{Bourgoin2015}
J.-P. Bourgoin, B.~L. Higgins, N.~Gigov, C.~Holloway, C.~J. Pugh, S.~Kaiser,
  M.~Cranmer, and T.~Jennewein, ``Free-space quantum key distribution to a
  moving receiver,'' {\em Optics Express}, vol.~23, p.~33437, 2015.

\bibitem{Pugh2017}
C.~J. Pugh, S.~Kaiser, J.-P. Bourgoin, J.~Jin, N.~Sultana, S.~Agne,
  E.~Anisimova, V.~Makarov, E.~Choi, B.~L. Higgins, and T.~Jennewein,
  ``Airborne demonstration of a quantum key distribution receiver payload,''
  {\em Quantum Sci. Technol.}, vol.~2, p.~024009, 2017.

\bibitem{Anisimova2017}
E.~Anisimova, B.~L. Higgins, J.-P. Bourgoin, M.~Cranmer, E.~Choi, D.~Hudson,
  L.~P. Piche, A.~Scott, V.~Makarov, and T.~Jennewein, ``Mitigating radiation
  damage of single photon detectors for space applications,'' {\em EPJ Quantum
  Technol.}, vol.~4, p.~10, 2017.

\bibitem{DSouza2021}
I.~DSouza, J.-P. Bourgoin, B.~L. Higgins, J.~G. Lim, R.~Tannous, S.~Agne,
  B.~Moffat, V.~Makarov, and T.~Jennewein, ``Repeated radiation damage and
  thermal annealing of avalanche photodiodes,'' {\em EPJ Quantum Technol.},
  vol.~8, p.~13, 2021.

\bibitem{Bourgoin2013}
J.-P. Bourgoin, E.~Meyer-Scott, B.~L. Higgins, B.~Helou, C.~Erven, H.~H\"ubel,
  B.~Kumar, D.~Hudson, I.~D'Souza, R.~Girard, R.~Laflamme, and T.~Jennewein,
  ``A comprehensive design and performance analysis of low earth orbit
  satellite quantum communication,'' {\em New J. Phys.}, vol.~15, p.~023006,
  2013.

\bibitem{Elvidge1999}
C.~D. Elvidge, K.~E. Baught, J.~B. Dietz, T.~Bland, P.~C. Sutton, and H.~W.
  Kroehl, ``Radiance calibration of {DMSP}-{OLS} low-light imaging data of
  human settlements,'' {\em Remote Sens. Environ.}, vol.~68, pp.~77--88, 1999.

\bibitem{Cinzano2001}
P.~Cinzano, F.~Falchi, and C.~D. Elvidge, ``The night sky in the world.''
  \url{http://www.lightpollution.it/dmsp/}, 2001.

\bibitem{berk2005modtran}
A.~Berk, G.~P. Anderson, P.~K. Acharya, L.~S. Bernstein, L.~Muratov, J.~Lee,
  M.~Fox, S.~M. Adler-Golden, J.~H. Chetwynd, M.~L. Hoke, {\em et~al.},
  ``Modtran 5: a reformulated atmospheric band model with auxiliary species and
  practical multiple scattering options: update,'' in {\em Algorithms and
  technologies for multispectral, hyperspectral, and ultraspectral imagery XI},
  vol.~5806, pp.~662--667, SPIE, 2005.

\bibitem{Erven2012}
C.~Erven, B.~Heim, E.~Meyer-Scott, J.-P. Bourgoin, R.~Laflamme, G.~Weihs, and
  T.~Jennewein, ``Studying free-space transmission statistics and improving
  free-space quantum key distribution in the turbulent atmosphere,'' {\em New.
  J. Phys.}, vol.~14, p.~123018, 2012.

\bibitem{acosta2021analysis}
V.~M. Acosta, D.~Dequal, M.~Schiavon, A.~Montmerle-Bonnefois, C.~B. Lim, J.-M.
  Conan, and E.~Diamanti, ``Analysis of satellite-to-ground quantum key
  distribution with adaptive optics,'' {\em arXiv preprint arXiv:2111.06747},
  2021.

\bibitem{roberts2018first}
L.~Roberts, G.~Block, S.~Fregoso, H.~Herzog, S.~Meeker, J.~Roberts, J.~Tesch,
  T.~Truong, J.~Rodriguez, and A.~Bechter, ``First results from the adaptive
  optics system for lcrd's optical ground station one,'' in {\em The Advanced
  Maui Optical and Space Surveillance Technologies Conference}, p.~3, 2018.

\bibitem{wright2015adaptive}
M.~W. Wright, J.~F. Morris, J.~M. Kovalik, K.~S. Andrews, M.~J. Abrahamson, and
  A.~Biswas, ``Adaptive optics correction into single mode fiber for a low
  earth orbiting space to ground optical communication link using the opals
  downlink,'' {\em Optics Express}, vol.~23, no.~26, pp.~33705--33712, 2015.

\bibitem{Pugh2020_AOT}
C.~J. Pugh, J.-F. Lavigne, J.-P. Bourgoin, B.~L. Higgins, and T.~Jennewein,
  ``Adaptive optics benefit for quantum key distribution uplink from ground to
  a satellite,'' {\em Advanced Optical Technologies}, vol.~9, no.~5,
  pp.~263--273, 2020.

\bibitem{scharlemann2011propulsion}
C.~Scharlemann, M.~Tajmar, I.~Vasiljevich, N.~Buldrini, D.~Krejci, and
  B.~Seifert, ``Propulsion for nanosatellites,'' in {\em 32nd International
  Electric Propulsion Conference, Wiesbaden, Germany, September 2011,
  IEPC-2011}, vol.~171, 2011.

\bibitem{Syarafana.2022}
S.~Shafeeq, ``Nanosatellite with singapore start-up's thruster deployed into
  space on spacex mission,'' {\em The Straits Times}.

\bibitem{Schwartz2016}
N.~Schwartz, D.~Pearson, S.~Todd, A.~Vick, D.~Lunney, and D.~MacLeod, ``A
  segmented deployable primary mirror for {Earth} observation from a cubesat
  platform,'' in {\em 30th annual AAIA/USU conference on small satellites
  (SSC16-WK-3)}, 2016.

\bibitem{Corbacho2020}
V.~V. Corbacho, H.~Kuiper, and E.~Gill, ``{Review on thermal and mechanical
  challenges in the development of deployable space optics},'' {\em Journal of
  Astronomical Telescopes, Instruments, and Systems}, vol.~6, no.~1, pp.~1 --
  30, 2020.

\bibitem{PhysRevA.93.010301}
D.~Dequal, G.~Vallone, D.~Bacco, S.~Gaiarin, V.~Luceri, G.~Bianco, and
  P.~Villoresi, ``Experimental single-photon exchange along a space link of
  7000 km,'' {\em Phys. Rev. A}, vol.~93, p.~010301, Jan 2016.

\bibitem{dirks2021geoqkd}
B.~Dirks, I.~Ferrario, A.~Le~Pera, D.~V. Finocchiaro, M.~Desmons, D.~de~Lange,
  H.~de~Man, A.~J. Meskers, J.~Morits, N.~M. Neumann, {\em et~al.}, ``Geoqkd:
  quantum key distribution from a geostationary satellite,'' in {\em
  International Conference on Space Optics—ICSO 2020}, vol.~11852,
  p.~118520J, International Society for Optics and Photonics, 2021.

\bibitem{ecker2021strategies}
S.~Ecker, B.~Liu, J.~Handsteiner, M.~Fink, D.~Rauch, F.~Steinlechner,
  T.~Scheidl, A.~Zeilinger, and R.~Ursin, ``Strategies for achieving high key
  rates in satellite-based qkd,'' {\em npj Quantum Information}, vol.~7, no.~1,
  pp.~1--7, 2021.

\bibitem{Holzman2019_AQT}
I.~Holzman and Y.~Ivry, ``Superconducting nanowires for single-photon
  detection: Progress, challenges, and opportunities,'' {\em Advanced Quantum
  Technologies}, vol.~2, no.~3-4, p.~1800058, 2019.

\bibitem{Lutkenhaus1999estimates}
N.~L\"utkenhaus, ``Estimates for practical quantum cryptography,'' {\em Phys.
  Rev. A}, vol.~59, pp.~3301--3319, May 1999.

\bibitem{Stahl2019Multivariable}
H.~P. Stahl, ``Multivariable {Parametric} {Cost} {Model} for {Ground} and
  {Space} {Telescope} {Assemblies},'' {\em Bulletin of the AAS}, vol.~51, sep
  30 2019.
\newblock https://baas.aas.org/pub/2020n7i143.

\bibitem{Nicolo2017_IOP}
N.~L. Piparo, N.~Sinclair, and M.~Razavi, ``Memory-assisted quantum key
  distribution resilient against multiple-excitation effects,'' {\em Quantum
  Science and Technology}, vol.~3, p.~014009, Dec 2017.

\bibitem{Sidhu2017_PRA}
J.~S. Sidhu and P.~Kok, ``Quantum metrology of spatial deformation using arrays
  of classical and quantum light emitters,'' {\em Phys. Rev. A}, vol.~95,
  p.~063829, June 2017.

\bibitem{Sidhu2018_arxiv}
J.~S. Sidhu and P.~Kok, ``Quantum {Fisher} information for general spatial
  deformations of quantum emitters,'' {\em ArXiv}, February 2018.

\bibitem{Sidhu2021_PRX}
J.~S. Sidhu, Y.~Ouyang, E.~T. Campbell, and P.~Kok, ``Tight bounds on the
  simultaneous estimation of incompatible parameters,'' {\em Phys. Rev. X},
  vol.~11, p.~011028, Feb 2021.

\bibitem{He2019}
M.~He, R.~Malaney, and J.~Green, ``Photonic engineering for cv-qkd over
  earth-satellite channels,'' in {\em ICC 2019 - 2019 IEEE International
  Conference on Communications (ICC)}, pp.~1--7, 2019.

\bibitem{Dequal2021_npjQI}
D.~Dequal, L.~Trigo~Vidarte, V.~Roman~Rodriguez, G.~Vallone, P.~Villoresi,
  A.~Leverrier, and E.~Diamanti, ``Feasibility of satellite-to-ground
  continuous-variable quantum key distribution,'' {\em npj Quantum
  Information}, vol.~7, p.~3, April 2021.

\bibitem{bartlett2007reference}
S.~D. Bartlett, T.~Rudolph, and R.~W. Spekkens, ``Reference frames,
  superselection rules, and quantum information,'' {\em Reviews of Modern
  Physics}, vol.~79, no.~2, p.~555, 2007.

\bibitem{laing2010reference}
A.~Laing, V.~Scarani, J.~G. Rarity, and J.~L. O’Brien,
  ``Reference-frame-independent quantum key distribution,'' {\em Physical
  Review A}, vol.~82, no.~1, p.~012304, 2010.

\bibitem{wang2015reference}
C.~Wang, S.-H. Sun, X.-C. Ma, G.-Z. Tang, and L.-M. Liang,
  ``Reference-frame-independent quantum key distribution with source flaws,''
  {\em Physical Review A}, vol.~92, no.~4, p.~042319, 2015.

\bibitem{tannous2019demonstration}
R.~Tannous, Z.~Ye, J.~Jin, K.~B. Kuntz, N.~L{\"u}tkenhaus, and T.~Jennewein,
  ``Demonstration of a 6 state-4 state reference frame independent channel for
  quantum key distribution,'' {\em Applied Physics Letters}, vol.~115, no.~21,
  p.~211103, 2019.

\bibitem{Jin2018_PRA}
J.~Jin, S.~Agne, J.-P. Bourgoin, Y.~Zhang, N.~L\"utkenhaus, and T.~Jennewein,
  ``Demonstration of analyzers for multimode photonic time-bin qubits,'' {\em
  Phys. Rev. A}, vol.~97, p.~043847, Apr 2018.

\bibitem{Jin2019_OE}
J.~Jin, J.-P. Bourgoin, R.~Tannous, S.~Agne, C.~J. Pugh, K.~B. Kuntz, B.~L.
  Higgins, and T.~Jennewein, ``Genuine time-bin-encoded quantum key
  distribution over a turbulent depolarizing free-space channel,'' {\em Opt.
  Express}, vol.~27, pp.~37214--37223, Dec 2019.

\bibitem{Tannous2023_arxiv}
R.~Tannous, W.~Wu, S.~Vinet, C.~Perumangatt, D.~Sinar, A.~Ling, and
  T.~Jennewein, ``Towards fully passive time-bin quantum key distribution over
  multi-mode channels,'' {\em ArXiv}, November 2023.

\bibitem{tittel2000}
W.~Tittel, J.~Brendel, H.~Zbinden, and N.~Gisin, ``Quantum cryptography using
  entangled photons in energy-time bell states,'' {\em Phys. Rev. Lett.},
  vol.~84, p.~4737, 2000.

\bibitem{islam2017}
N.~T. Islam, C.~C.~W. Lim, C.~Cahall, J.~Kim, and D.~J. Gauthier, ``Provably
  secure and high-rate quantum key distribution with time-bin qudits,'' {\em
  Science advances}, vol.~3, p.~e1701491, 2017.

\bibitem{frameqkd2016}
T.~Brougham, C.~F. Wildfeuer, S.~M. Barnett, and D.~J. Gauthier, ``The
  information of high-dimensional time-bin encoded photons,'' {\em Eur. Phys.
  J. D.}, vol.~70, p.~214, 2016.

\bibitem{castillo2020}
A.~Tello~Castillo, C.~Novo, and R.~Donaldson, ``Prospects of time-bin quantum
  key distribution in turbulent free-space channels,'' {\em Emerging Imaging
  and Sensing Technologies for Security and Defence V; and Advanced
  Manufacturing Technologies for Micro- and Nanosystems in Security and Defence
  III, SPIE}, vol.~11540, p.~1154006, 2020.

\bibitem{alleaume2004experimental}
R.~All{\'e}aume, F.~Treussart, G.~Messin, Y.~Dumeige, J.-F. Roch, A.~Beveratos,
  R.~Brouri-Tualle, J.-P. Poizat, and P.~Grangier, ``Experimental open-air
  quantum key distribution with a single-photon source,'' {\em New Journal of
  physics}, vol.~6, no.~1, p.~92, 2004.

\bibitem{takemoto2015quantum}
K.~Takemoto, Y.~Nambu, T.~Miyazawa, Y.~Sakuma, T.~Yamamoto, S.~Yorozu, and
  Y.~Arakawa, ``Quantum key distribution over 120 km using ultrahigh purity
  single-photon source and superconducting single-photon detectors,'' {\em
  Scientific reports}, vol.~5, no.~1, p.~14383, 2015.

\bibitem{sangouard2012single}
N.~Sangouard and H.~Zbinden, ``What are single photons good for?,'' {\em
  Journal of Modern Optics}, vol.~59, no.~17, pp.~1458--1464, 2012.

\bibitem{Lvovsky2009_NP}
A.~I. Lvovsky, B.~C. Sanders, and W.~Tittel, ``Optical quantum memory,'' {\em
  Nature Photonics}, vol.~3, pp.~706--714, December 2009.

\bibitem{muller2014demand}
M.~M{\"u}ller, S.~Bounouar, K.~D. J{\"o}ns, M.~Gl{\"a}ssl, and P.~Michler,
  ``On-demand generation of indistinguishable polarization-entangled photon
  pairs,'' {\em Nature Photonics}, vol.~8, no.~3, pp.~224--228, 2014.

\bibitem{naughton2019design}
D.~P. Naughton, R.~Bedington, S.~Barraclough, T.~Islam, D.~Griffin, B.~Smith,
  J.~Kurtz, A.~S. Alenin, I.~J. Vaughn, A.~Ramana, {\em et~al.}, ``Design
  considerations for an optical link supporting intersatellite quantum key
  distribution,'' {\em Optical Engineering}, vol.~58, no.~1, p.~016106, 2019.

\bibitem{islam2017approaches}
T.~Islam, R.~Bedington, and A.~Ling, ``Approaches to a global quantum key
  distribution network,'' in {\em Quantum Information Science and Technology
  III}, vol.~10442, pp.~28--33, SPIE, 2017.

\bibitem{dequal2021feasibility}
D.~Dequal, L.~Trigo~Vidarte, V.~Roman~Rodriguez, G.~Vallone, P.~Villoresi,
  A.~Leverrier, and E.~Diamanti, ``Feasibility of satellite-to-ground
  continuous-variable quantum key distribution,'' {\em npj Quantum
  Information}, vol.~7, no.~1, pp.~1--10, 2021.

\end{thebibliography}

\end{document}